\documentclass[12pt,superscriptaddress]{revtex4}
\usepackage{latexsym, color, graphicx}
\usepackage{amssymb}
\usepackage{amsmath}

\newcommand{\vect}[1]{\mbox{\boldmath $#1$}}

\newcommand{\omegase}{\omega_{*}}
\newcommand{\nuee}{\nu_{ee}}
\newcommand{\energy}{\mathcal{E}}
\newcommand{\sgn}{\mathrm{sgn}}

\newcommand{\transpose}{\mathrm{T}}

\begin{document}

\title{Generalized universal instability: Transient linear amplification and subcritical turbulence}



\author{Matt Landreman}
\email[]{mattland@umd.edu}
\affiliation{Institute for Research in Electronics and Applied Physics, University of Maryland, College Park, MD, 20742, USA}
\author{Gabriel G. Plunk}
\affiliation{Max Planck Institute for Plasma Physics, EURATOM Association and Max-Planck/Princeton Research Center for Plasma Physics}
\author{William Dorland}
\affiliation{Institute for Research in Electronics and Applied Physics, University of Maryland, College Park, MD, 20742, USA}


\date{\today}

\begin{abstract}

In this work we numerically demonstrate both significant transient (i.e. non-modal) linear amplification
and sustained nonlinear turbulence
in a kinetic plasma system with no unstable eigenmodes.
The particular system considered is an electrostatic plasma slab
with magnetic shear, kinetic electrons and ions,
weak collisions,
and a density gradient, but with no temperature gradient.
In contrast to hydrodynamic examples of non-modal growth and subcritical turbulence,
here there is no sheared flow in the equilibrium.
Significant transient linear amplification
is found when the magnetic shear and collisionality are weak.
It is also demonstrated that nonlinear turbulence can be sustained if initialized
at sufficient amplitude.
We prove these two phenomena are related:
when sustained turbulence occurs without unstable eigenmodes,
states that are typical of the turbulence must
yield transient linear amplification of the gyrokinetic free energy.

\end{abstract}


\maketitle 

\section{Introduction}

Despite the usefulness of linear eigenmode analysis in many contexts,
it has been established that
such analysis can in practice give misleading predictions
about the stability of various systems \cite{TrefethenEmbree}.
In particular, turbulence may be sustained in a number of fluid systems
despite the absence of any unstable eigenmodes.
The first such systems known were certain sheared flow patterns in
neutral fluids, such as plane Poiseuille or plane Couette flow
\cite{Tillmark,TrefethenSubcritical,Grossman}.
Later, sustained turbulence was seen
in fluid simulations of several plasma systems with no absolute linear instabilities
\cite{Waltz, Scott1, Scott2, Drake}.
More recently,
turbulence in the absence of linear instability has been seen in kinetic plasma simulations involving
sheared equilibrium flow \cite{EdmundPRL, BarnesRotation}.
These examples demonstrate that linear eigenmode analysis can give
deceiving predictions for nonlinear behavior,
a conclusion supported by other studies of plasma turbulence in which
linear instability exists, but
the fluctuations in saturated nonlinear turbulence are different in character from the linear
eigenmodes \cite{Biskamp,Friedman1,Friedman2}.

For the cases of turbulence without absolute linear stability in the presence of sheared flows,
for both neutral fluids and plasmas, the phenomenon of transient linear amplification
appears to be important
\cite{TrefethenSubcritical, FarrellIoannou, Grossman, DelSoleSurvey, TrefethenEmbree, Schmid, Newton, Alex}.
Also known as non-modal amplification, transient amplification is a period of growth before
the exponential decay sets in.
A necessary (but not sufficient) condition for transient amplification
is that the linear operator must not have a complete orthogonal set of eigenvectors.
Analysis of transient amplification has also been applied to plasma systems in which
linear instability exists but transient linear behavior may be
more important \cite{Camargo, Camporeale2010, Friedman2, SquirePRL, SquireApJ}.

In the present work, we consider the phenomena of transient amplification and turbulence
without instability for the case of the weakly-collisional
plasma drift wave driven by a density gradient\cite{Galeev,Krall3},
neglecting toroidal effects.
This oscillation has been called the ``universal instability'' as it was at first
mistakenly thought to be linearly unstable in any plasma with a density gradient.
The oscillation is electrostatic in nature, and linear instability can indeed be found
in slab geometry without magnetic shear or temperature gradients.
It was later established that when any amount of magnetic shear is included in the model,
no matter how small,
the eigenmodes become linearly damped for $k_y \rho_i \ll 1$,
where $k_y$ is the perpendicular wavenumber and $\rho_i$ is the ion gyroradius
\cite{Ross, Tsang, Antonsen}.
However, instability can persist at $k_y \rho_i \ge 1$ when the collisionality is
very low \cite{usUniversalInstability}.
When both magnetic shear and modest collisions are included \cite{usUniversalInstability},
the mode becomes linearly stable at all $k_y \rho_i$.
It is this latter situation that we will consider here.
(As an aside, the small-but-finite-shear eigenmodes
are different from zero-shear modes not only for the
universal instability, but also for the slab ion-temperature-gradient (ITG) instability.
Specifically, in \cite{Plunk2014}, considering the slab mode by setting
curvature drift effects $\omega_d$ to 0, the small-shear limit of (19) is different from
the zero shear result (12).)

The fluid version of this drift wave system was investigated in \cite{Drake},
using a model with three fluctuating quantities (density, electrostatic potential, and parallel velocity).
In this model all linear modes were damped, yet nonlinearly, turbulence could be sustained.
It was pointed out that an important secondary
instability in this system is a drift wave rotated by $\pi/2$ in the plane
perpendicular to the magnetic field, since the stabilizing effects of magnetic
shear are reduced in this orientation.
In this rotated drift instability, the role of the equilibrium density gradient is replaced in a sense by the
perturbed density gradient, the latter represented in the nonlinear term.
However, since the nonlinearity is conservative, the rotated drift wave does not inject energy
into the system in the way that can be done by a perturbation feeding off the density gradient of the equilibrium.
Several authors have also investigated transient linear effects in fluid drift wave models
\cite{Camargo, Friedman1, Friedman2}. However, these studies did not
include magnetic shear, which we will find to have a very strong effect on
transient behavior.

In the following sections, we establish three results concerning the drift wave system in slab geometry
with magnetic shear,
extending previous work by considering a kinetic description and exploring the role of transient linear amplification.
For these studies, the initial-value gyrokinetic code gs2 \cite{gs2} will be used
for numerical computations.
First, in section \ref{sec:transient} we demonstrate that
significant transient linear amplification is possible in this system,
and that the transiently growing fluctuations
resemble the zero-shear eigenmodes
in the dependence of their instantaneous growth rate on wavenumber.
Second, in section \ref{sec:turbulence} we demonstrate that universal-mode turbulence without linear
instability, seen previously in the fluid model in \cite{Drake},
is also possible in the gyrokinetic model.
Finally, we prove there is a relationship between
transient linear amplification and nonlinear turbulence in this system,
using similar arguments to the proof for hydrodynamics in \cite{DelSoleNecessity}
and arriving at similar conclusions.
Specifically, if the physical parameters are such that turbulence can be sustained,
typical states of the turbulence must be amplified linearly for some amount of time, even if these states
eventually decay exponentially.
Two more precise statements of this result and their proof will be given in section \ref{sec:proof}.

\section{Gyrokinetic model}
\label{sec:model}

We consider slab geometry with magnetic shear, so the magnetic field in Cartesian
coordinates $(x',y',z')$ is $\vect{B} = [\vect{e}_{z'} + (x' / L_s) \vect{e}_{y'} ]B$
for some constant $B$ and shear length $L_s$.  A field-aligned coordinate system
is introduced: $x = x', \;y = y' - x' z' / L_s, \; z = z'$,
so $\vect{B}\cdot\nabla x = \vect{B}\cdot\nabla y = 0$ and $\vect{B}\cdot\nabla z = B$.
In this geometry, we consider the nonlinear electrostatic
gyrokinetic equation \cite{FriemanChen},
dropping toroidal effects:
\begin{eqnarray}
\label{eq:gke}
\frac{\partial h_s}{\partial t}
+ v_{||} \frac{\partial h_s}{\partial z}
+ \frac{1}{B} \left[
\frac{\partial \left<\Phi\right>_{\vect{R}_s}}{\partial X_s} \frac{\partial h_s}{\partial Y_s}
-\frac{\partial \left<\Phi\right>_{\vect{R}_s}}{\partial Y_s} \frac{\partial h_s}{\partial X_s}
\right]
-\left< C_s\left\{h_s\right\} \right>_{\vect{R}_s} \\
=
\frac{q_s}{T_s} f_{Ms} \frac{\partial \left<\Phi\right>_{\vect{R}_s}}{\partial t}
-\frac{f_{Ms}}{B L_n} \frac{\partial \left< \Phi\right>_{\vect{R}_s}}{\partial Y_s}
\left[ 1 + \left( \energy_s-\frac{3}{2}\right) \eta_s\right],
\nonumber
\end{eqnarray}
together with the quasineutrality condition
\begin{equation}
-\frac{n e \Phi}{T_i} + \int d^3v \left<h_i\right>_{\vect{r}}
=
\frac{n e \Phi}{T_e} + \int d^3v \left<h_e\right>_{\vect{r}}.
\label{eq:qn}
\end{equation}
Here $s \in \left\{i,e\right\}$ subscripts denote species,
$\Phi(t,\vect{r})$ is the electrostatic potential, $h_s(t,\vect{R}_s,\energy_s,\mu)$ is the nonadiabatic distribution function,
$\energy_s = m_s v^2/(2 T_s)$, $\mu = v_{\bot}^2/(2B)$,
and $f_{Ms}$ is the leading-order Maxwellian.
The scale length of the equilibrium density $n$ is $L_n = -n/(dn/dx)$,
$q_i = e = -q_e$ is the proton charge,
$T_s$ is the species temperature,
$C_s$ is the collision operator,
and
$\left<\right>_{\vect{r}}$
and
$\left<\right>_{\vect{R}_s}$
denote gyroaverages at fixed particle position $\vect{r}=(x,y,z)$ and guiding-center position $\vect{R}_s=(X_s,Y_s,Z_s)$.
The distinction between $z$ and $Z_s$ may be neglected.
Expanding the fluctuating quantities in spatial Fourier modes as $\Phi = \sum_{k_x,k_y} \Phi_{k_x,k_y}(t,z) \exp(i k_x x + i k_y y)$
and $h_s = \sum_{k_x,k_y} h_{s,k_x,k_y}(t,z,\energy_s,\mu) \exp(i k_x X_s + i k_y Y_s)$,
we may replace the gyro-averages with Bessel functions:
$\left<\Phi_{k_x,k_y} \exp(i k_x x + i k_y y)\right>_{\vect{R}_s}
=J_{0s} \Phi_{k_x,k_y} \exp(i k_x X_s + i k_y Y_s)$
and
$\left<h_{s,k_x,k_y}\exp(i k_x X_s + i k_y Y_s)\right>_{\vect{r}} = J_{0s} h_{s,k_x,k_y}\exp(i k_x x + i k_y y)$
with $J_{0s} = J_0(k_{\perp} v_{\perp} / \Omega_s)$
and $\Omega_s = e B/m_s$.
The perpendicular wavenumber is given by
\begin{equation}
k_{\perp} = \sqrt{k_y^2 + (k_x - k_y z/L_s)^2}.
\label{eq:kperp}
\end{equation}
The system is uniquely specified by the following dimensionless parameters:
$L_s / L_n$, $\eta_i$, $\eta_e$, $T_e / T_i$, and collisionality $\nuee L_n / v_i$,
where $\nuee = \sqrt{2}\pi n e^4 \ln\Lambda/[(4\pi\epsilon_0)^2 m_e^{1/2} T_e^{3/2}]$ is the electron-electron collision frequency
and $v_i = \sqrt{2 T_i / m_i}$.

Notice that the parallel coordinate $z$ enters the system (\ref{eq:gke})-(\ref{eq:kperp}) only through (\ref{eq:kperp}).
Also notice $k_x$ enters the system (\ref{eq:gke})-(\ref{eq:kperp}) only through (\ref{eq:kperp}) and through the nonlinear term in (\ref{eq:gke}).
Therefore, a linear mode at nonzero $k_x$ and nonzero $k_y$ satisfies exactly the same equations as the corresponding $k_x=0$ mode, but translated in $z$ by
$L_s k_x / k_y$.  As a consequence, the linear growth or damping rates of $k_y \ne 0$ modes
are independent of $k_x$, and the eigenmode structures are independent of $k_x$
up to this translation in $z$.

One important property of the system (\ref{eq:gke})-(\ref{eq:qn})
is the free energy conservation equation \cite{Howes, AbelReview}
obtained from the $\sum_s \int d^3R_s \int d^3v (T_s h_s/f_{Ms}) (\ldots)$ moment of (\ref{eq:gke}):
\begin{equation}
\frac{d W}{d t} = S(h_i,h_e),
\label{eq:dWdt}
\end{equation}
where
\begin{equation}
\label{eq:W}
W = \sum_s \int d^3r \int d^3v \frac{T_s \, \delta\! f_s^2}{2 f_{Ms}}
\end{equation}
is the free energy, $\delta\! f_s = h_s - q_s \Phi f_{Ms}/T_s$ is the total departure of the distribution
function from a Maxwellian,
and
\begin{eqnarray}
\label{eq:freeEnergySources}
S(h_i,h_e)
&=&
\sum_s \int d^3r \int d^3v \frac{T_s h_s}{f_{Ms}}  C_s\{ h_s \}  \\
&&+\sum_s
\int d^3R_s \int d^3v \frac{h_s}{B} \frac{\partial \left< \Phi\right>_{\vect{R}_s}}{\partial Y_s}
 \frac{T_s}{n_s}\frac{dn_s}{dx}
\left[ 1 + \left( \energy_s-\frac{3}{2}\right) \eta_s\right]
\nonumber
\end{eqnarray}
is the net energy source.
(To derive (\ref{eq:dWdt})-(\ref{eq:freeEnergySources}), the equalities
$\int d^3 R_s \int d^3v \left< \ldots\right>_{\vect{R}_s}
= \int d^3 R_s \int d^3v (\ldots)
\approx \int d^3 r \int d^3v (\ldots)
= \int d^3 r \int d^3v \left< \ldots\right>_{\vect{r}}
$ are used.)
The first term on the right-hand side of (\ref{eq:freeEnergySources})
is negative-definite (by the $H$-theorem).
The second term has the form of the particle and heat fluxes multiplied by
the density and temperature gradients, and this term increases $W$ if the fluxes have the appropriate sign
to relax the gradients.
Thus, in a turbulent steady state where $W$ is statistically constant,
equations (\ref{eq:dWdt})-(\ref{eq:freeEnergySources}) indicate
that energy is injected by the equilibrium gradient(s)
and dissipated by collisions.
Note that the operation $\int d^3R_s$ used in the derivation of
(\ref{eq:dWdt}) annihilates the nonlinear term
in (\ref{eq:gke}), and so (\ref{eq:dWdt})-(\ref{eq:freeEnergySources})
apply to both the linear and nonlinear dynamics.

The system (\ref{eq:gke})-(\ref{eq:qn}) may or may not possess absolute linear instability,
depending on the physical parameters.
For the rest of this paper, we take $\eta_i = \eta_e = 0$
so it is certain that the ion-temperature-gradient and electron-temperature-gradient (ETG)
instabilities are suppressed.
Even without temperature gradients, the system may still be unstable to the universal instability
driven by the density gradient.
For reference, the growth rates and frequencies of this instability
in the absence of magnetic shear are given in Appendix \ref{appendix:zeroShear}.
As detailed in the appendix, some insight into the instability can be obtained by expanding
the plasma dispersion function $Z$ for the electrons in the small-argument limit,
$Z(\omega / (|k_{||}| v_e)) \approx i \sqrt{\pi}$ for $|\omega / (k_{||} v_e)| \ll 1$,
and by expanding the ion $Z$ function in the large argument limit,
$Z(\omega / (|k_{||}| v_i)) \approx - |k_{||}| v_i / \omega$ for $|\omega / (k_{||} v_i)| \gg 1$.
One then finds a frequency with positive imaginary part $\propto i \sqrt{\pi} / L_n$,
indicating the instability is due to resonant electrons in the presence of a density gradient.
When an arbitrarily small amount of magnetic shear is introduced, the global eigenmodes are
no longer recognizable as the unsheared slab modes, and indeed are absolutely stable
for $k_y\rho_i \ll 1$ \cite{Ross,Tsang}.  One might reasonably expect weakly sheared systems with
stable global eigenmodes to exhibit transient linear amplification, with the number of $e$-foldings
of amplification tending to infinity as the magnetic shear tends to zero.  In the following section,
we will show this hypothesis to be true, and thereby demonstrate the apparently singular limit of zero shear
to be a continuous limit of the general shear case.

\section{Transient linear amplification}
\label{sec:transient}

\begin{figure}[h!]
\includegraphics[width=3in]{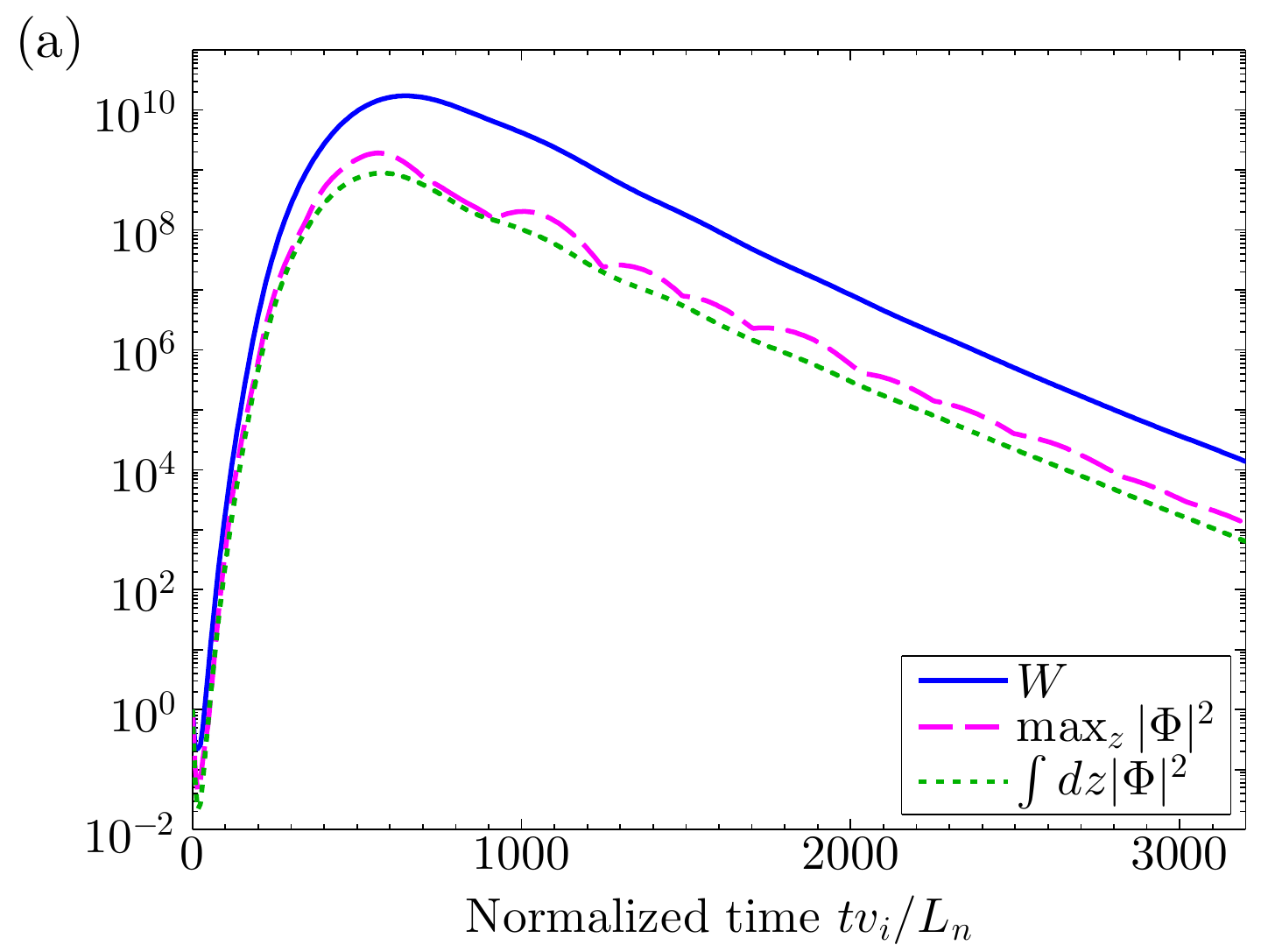}
\includegraphics[width=3in]{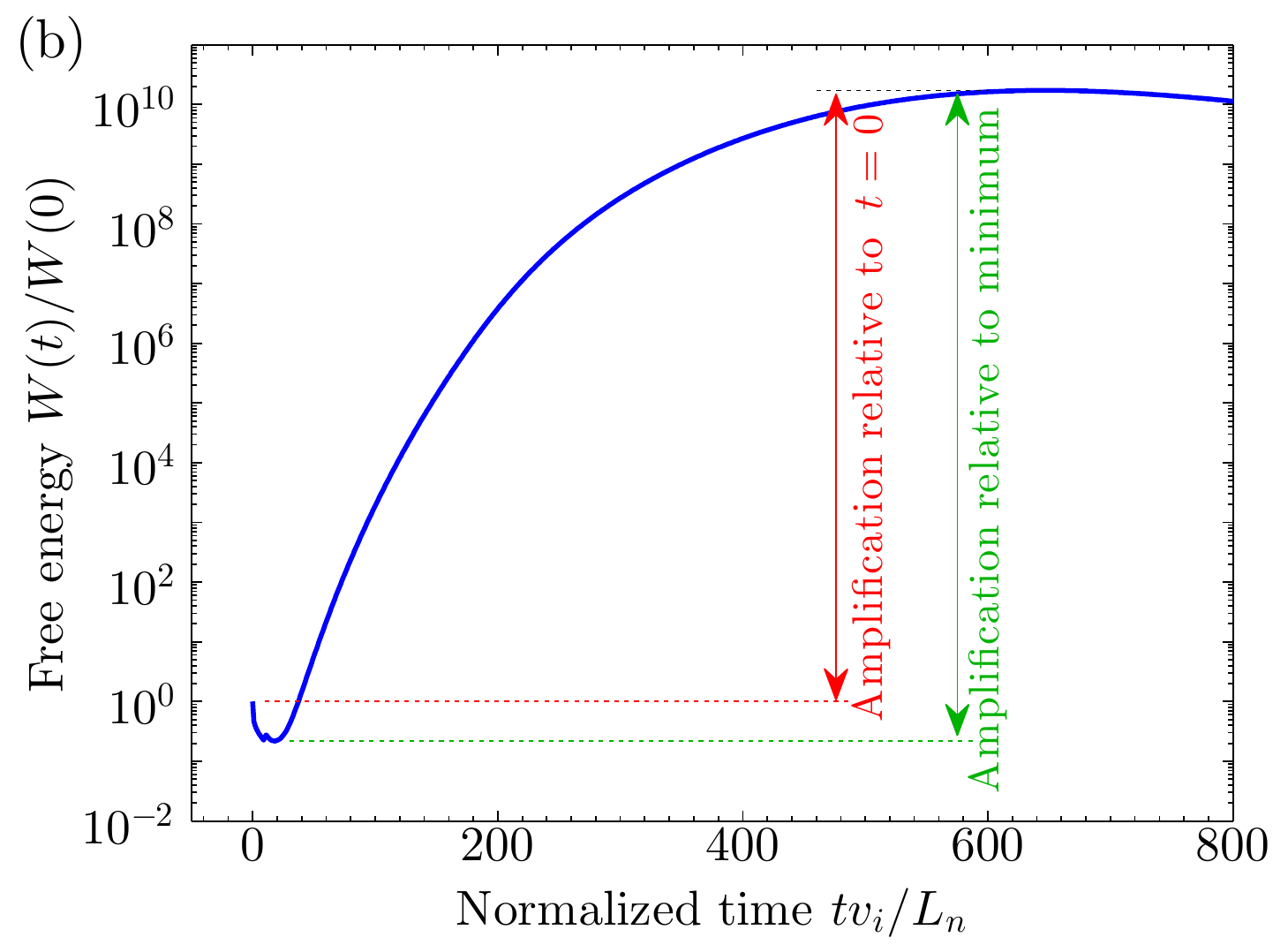}
\includegraphics[width=3in]{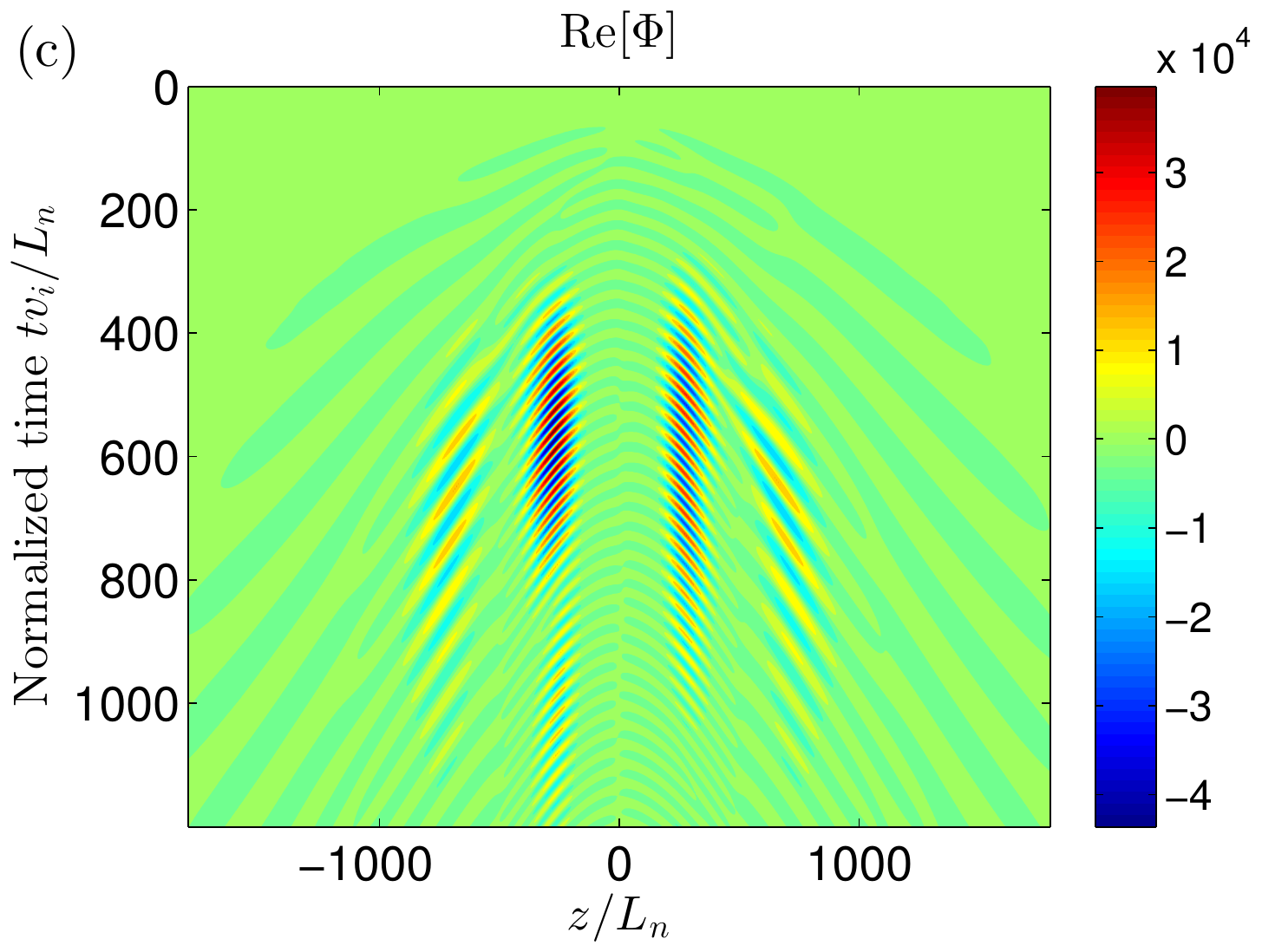}
\caption{(Color online) (a) Linear evolution for $L_s/L_n=300$, $\eta_i = \eta_e = 0$,
$\nuee = 0.05 v_i/L_n$, and $k_y \rho_i = 1$, illustrating transient amplification in three norms.
The initial conditions are detailed in the main text.
(b) The amount of transient amplification may be defined relative to $t=0$ or relative
to the minimum that occurs shortly thereafter.
(c) The potential structure along a field line for this example.
\label{fig:transientAmplification}}
\end{figure}

Turning now to linear initial-value simulations,
we indeed find that significant transient linear amplification is possible for certain parameters.
Figure \ref{fig:transientAmplification}.a illustrates
a linear computation for a single $(k_x,k_y)$ in a case of very weak magnetic shear, $L_s / L_n = 300$.
Other parameters used are $k_y \rho_i = 1$ (where  $\rho_i = v_i / \Omega_i$), $\nuee L_n / v_i = 0.05$,
$m_i = 3672 m_e$,
$T_e = T_i$, and $\eta_i = \eta_e = 0$.
The results are independent of $k_x$ up to a translation in $z$, as noted previously.
Collisions are also included to eliminate absolute instability for $k_y \rho_i \ge 0.7$,
as discussed in \cite{usUniversalInstability},
using the collision operators detailed in Ref. \cite{Abel}.
As transient amplification generally depends on the choice
of norm, here we compare three reasonable quadratic norms: $W$, $\int dz |\Phi|^2$,
and the maximum of $|\Phi|^2$ along $z$.
The curve for each norm is scaled by a constant so the initial amplitude is 1.
Amplification by several orders of magnitude is observed in all norms
before exponential decay sets in.

Results for the three norms shown are quite similar,
though the curve for the $W$ norm appears slightly higher than the other curves
due to less decay in the initial period $t v_i / L_n <20$.
Indeed, we find
all results described or plotted for the remainder of this paper are quite similar
for all of these norms.
Henceforth we restrict our attention to the $W$ norm since it is well motivated
by the conservation equation (\ref{eq:dWdt}).

For the calculation in figure \ref{fig:transientAmplification}, the initial condition used
is the standard ``noise'' initial condition of gs2:
$g_s = h_s - \left< \Phi\right>_{\vect{R}_s} q_s f_{Ms} / T_s$ is set to a Maxwellian velocity distribution
times a random complex number at each grid point in $z$.
(The real and imaginary parts of these complex numbers are each taken from a uniform distribution on $[-1,1]$.)
While techniques exist to calculate the optimal perturbation
that gives maximum amplification \cite{FarrellIoannou, SquirePRL, SquireApJ}, such techniques
are numerically challenging for this kinetic model due to the high dimensionality
of the problem.
The noise initial condition is a convenient shortcut, ensuring energy is likely
to be given to a growing perturbation if one exists.
It has been shown in \cite{Camargo, SquireApJ} that for studies of transient linear
amplification, random initial conditions are a reasonable proxy
for a directly calculated optimal initial condition.

Figure \ref{fig:transientAmplification}.b shows the same computation, zooming in to the initial transient period.
Before the amplification begins around $t v_i/L_n=20$, there is an initial decay period due to strongly
damped modes. An amplification factor may be defined relative to either to this minimum amplitude,
or relative to the amplitude at $t=0$.
In both cases, we define the amplification factor to be 1
if there is no such transient amplification.

\begin{figure}[h!]
\includegraphics[width=3in]{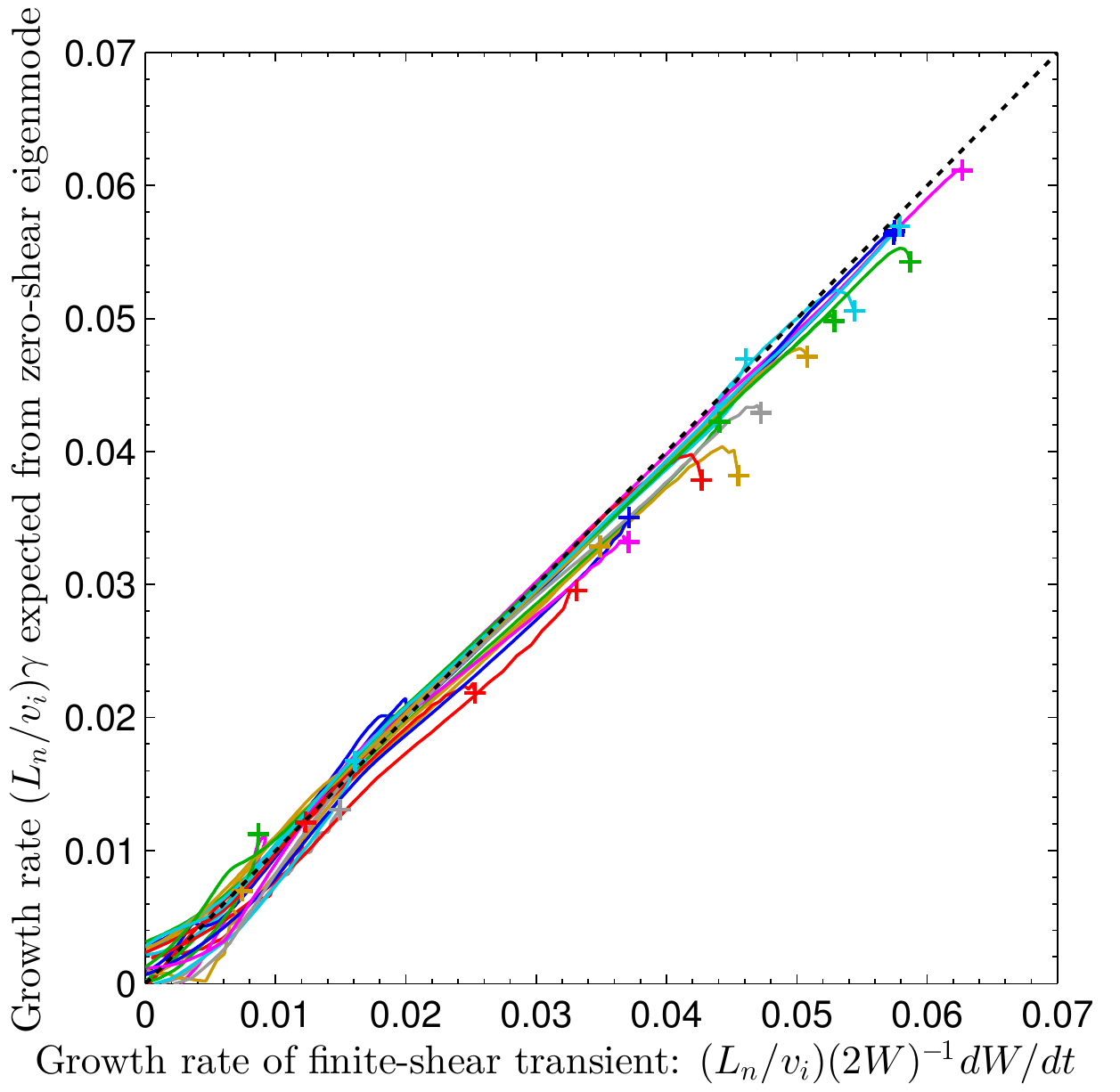}
\caption{(Color online)
Although a change from zero to small finite magnetic shear causes a fundamental change in the eigenvalues,
(which change from unstable to decaying),
the \emph{transient} finite-shear structures remain similar to the zero-shear modes.  For as shown here,
the instantaneous growth rate of the transient finite-shear structure is nearly identical
to the growth rate of a zero-shear eigenmode at the dominant $k_{\perp}$ and $k_{||}$ values
(\ref{eq:expectedKperp})-(\ref{eq:expectedKpar}).  Each trajectory shown is an independent
linear initial-value calculation, with $L_s/L_n$ varying from 20-300 between simulations.  The growth rate is plotted from
the time of maximum growth rate (at the $+$ symbols) until decay begins.
\label{fig:comparingTransientTo0Shear}}
\end{figure}

The transiently growing structures in the presence of magnetic shear
are closely related to the zero-shear eigenmodes.  To demonstrate this relationship, we consider a single
$(k_x,k_y)$, and define an effective
$k_{\perp}$ at each time for the transient fluctuation by
\begin{equation}
\label{eq:expectedKperp}
\left< k_{\perp} \right>(t) = \frac{\int dz\; k_{\perp}(z) \left| \Phi_{k_x,k_y}(t,z)\right|^2}
{\int dz\; k_{\perp}(z) \left| \Phi_{k_x,k_y}(t,z)\right|^2}.
\end{equation}
Similarly, we define an effective $|k_{||}|$ at each time for the transient structure
by
\begin{equation}
\label{eq:expectedKpar}
\left< \left| k_{||} \right| \right>(t) = \frac{\int dk_{||}  \left| k_{||}\right| \left| \tilde\Phi_{k_x,k_y}(t,k_{||})\right|^2}
{\int dk_{||}\left| \tilde\Phi_{k_x,k_y}(t,k_{||})\right|^2},
\end{equation}
where $\tilde\Phi_{k_x,k_y}$ is the Fourier transform of $\Phi_{k_x,k_y}$ in $z$. Using these effective values
of $k_{\perp}$ and $k_{||}$, the zero-shear dispersion relation may be evaluated.
The resulting growth rate can then be compared to the instantaneous growth rate of the finite-shear
transient structure, $\gamma = (2W(t))^{-1} dW/dt$.  This comparison is shown in figure \ref{fig:comparingTransientTo0Shear}.
In this figure, we display 25 independent simulations of transients,
varying $L_s/L_n$ between 20 and 300, and plotting the evolution of $\gamma$ from the time
of peak growth rate to the time at which the energy begins to decay.
In each case, there is very close agreement between the instantaneous growth rate of the transient
and the `predicted' growth rate from the zero-shear dispersion relation.
Thus, we can conclude that the transiently growing structures are closely related to the zero-shear
eigenmodes.  Magnetic shear causes $k_{\perp}$ and $k_{||}$ to vary with time as the structure
evolves (unlike the zero-shear case in which $k_{\perp}$ and $k_{||}$ are constants.)
However, at each instant, the finite-shear transient grows at precisely the rate expected from the zero-shear
instability at the average wavenumbers $\left< k_{\perp} \right>$ and $\left< \left| k_{||} \right| \right>$.
For the simulations in figure  \ref{fig:comparingTransientTo0Shear}, $k_y \rho_i = 2$ and $\nuee = 0.05 v_i/L_n$.

For the range of $k_{||}$ and $k_{\perp}$ values computed for the transient structures
of figure \ref{fig:comparingTransientTo0Shear}, the zero-shear dispersion relation
is well approximated by the approximate analytic dispersion relation (\ref{eq:approxRoot}).  Thus, we can also conclude
that the transient structures, like the zero-shear eigenmodes,
are driven by the combination of resonant electrons and the density gradient.

\begin{figure}[h!]
\includegraphics[width=3in]{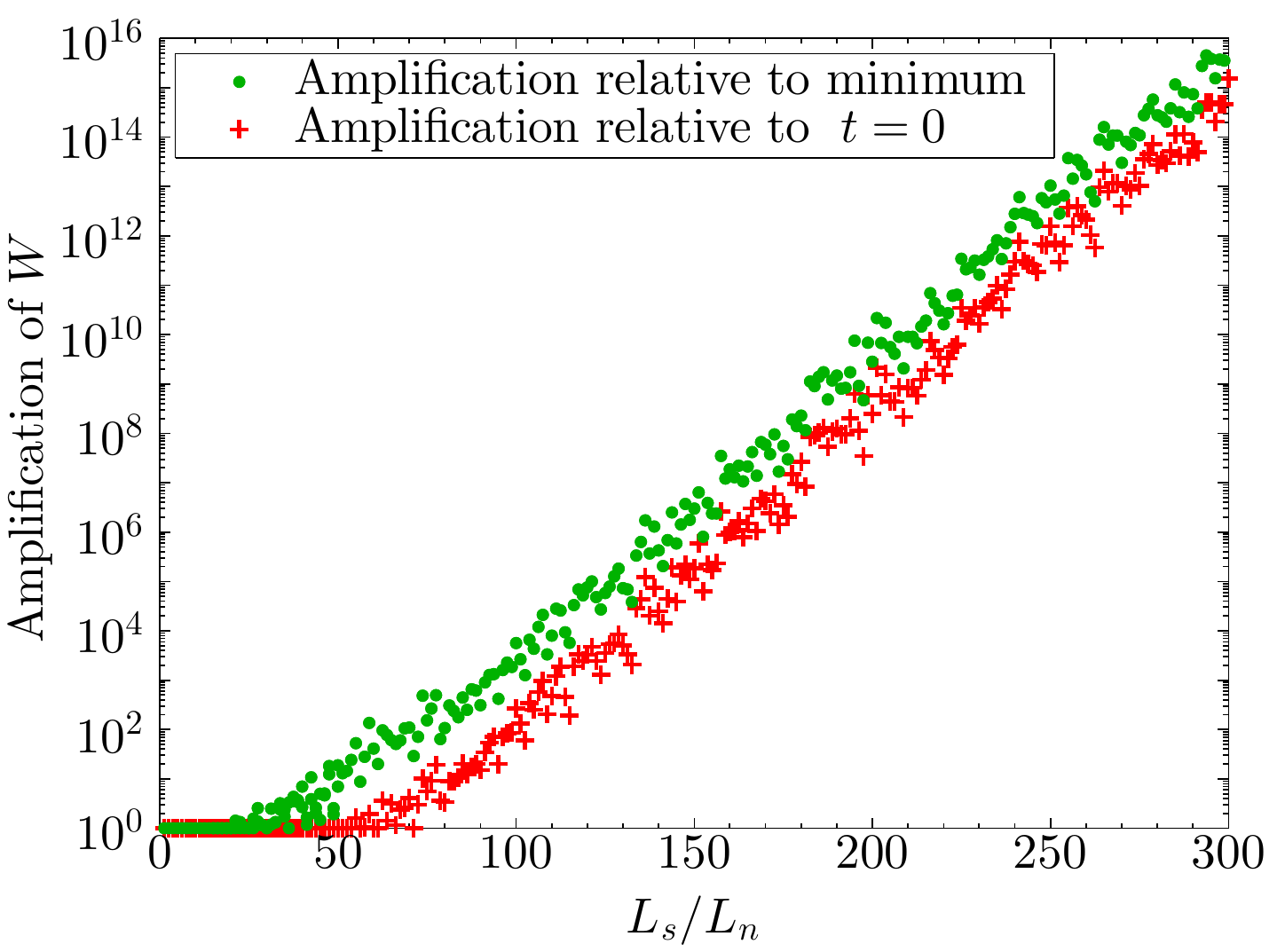}
\includegraphics[width=3in]{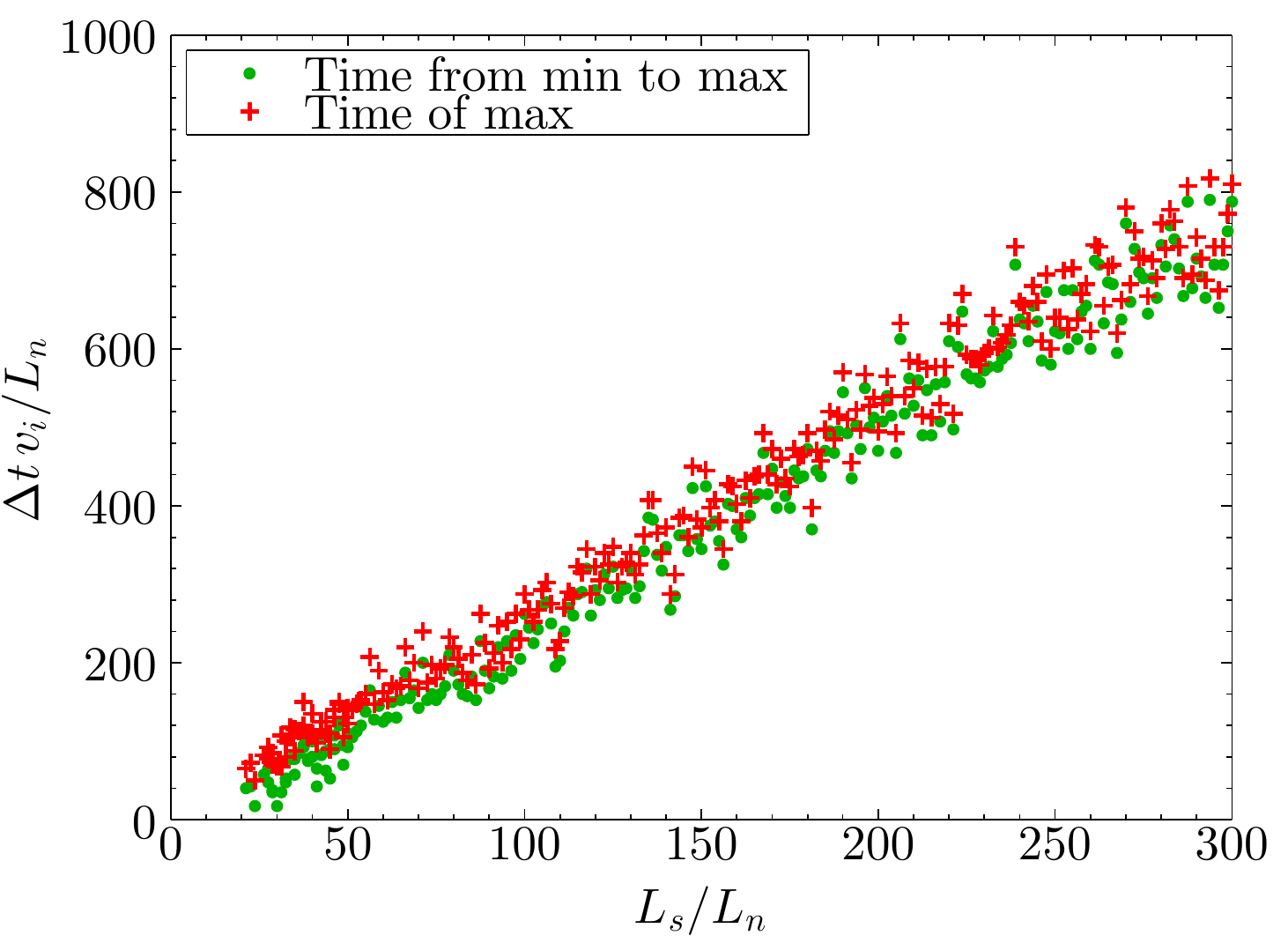}
\caption{(Color online) The (a) number of $e$-foldings of transient amplification and (b)
duration of the amplification both scale with the magnetic shear length scale $L_s$,
consistent with the appearance of absolute instability in the limit of vanishing shear ($L_s \to \infty$).
Parameters used are $k_y \rho_i = 2$ and $\nuee = 0.05 v_i / L_n$.
\label{fig:transientAmplificationVsLsOverLn}}
\end{figure}

Transient amplification is exponentially reduced by increasing magnetic shear,
as illustrated in figure \ref{fig:transientAmplificationVsLsOverLn}.
For this figure, $k_y \rho_i$ is fixed at 2, and other parameters
are the same as in figure \ref{fig:transientAmplification}.
(We choose $k_y \rho_i=2$ here since
wavenumbers near this value will turn out to be the most
amplified, as we will find shortly.)
There is some scatter in figure \ref{fig:transientAmplificationVsLsOverLn}, associated with the random initial condition,
but the exponential trend with $L_s/L_n$ is clear.
Notice that amplification only appears possible if $L_s/L_n \gtrsim 20$.
Values of $L_s/L_n$ far in excess of 20 are obtained in the edge of tokamak experiments.
(For a high aspect ratio tokamak, $L_s \sim qR/\hat{s}$ where $q$ is the safety factor, $R$ is
the major radius, $\hat{s} = (r/q) dq/dr$, and $r$ is the minor radius.)
Figure \ref{fig:transientAmplificationVsLsOverLn}.b demonstrates that the time during which amplification occurs
varies approximately linearly with $L_s/L_n$.

\begin{figure}[h!]
\includegraphics[width=3in]{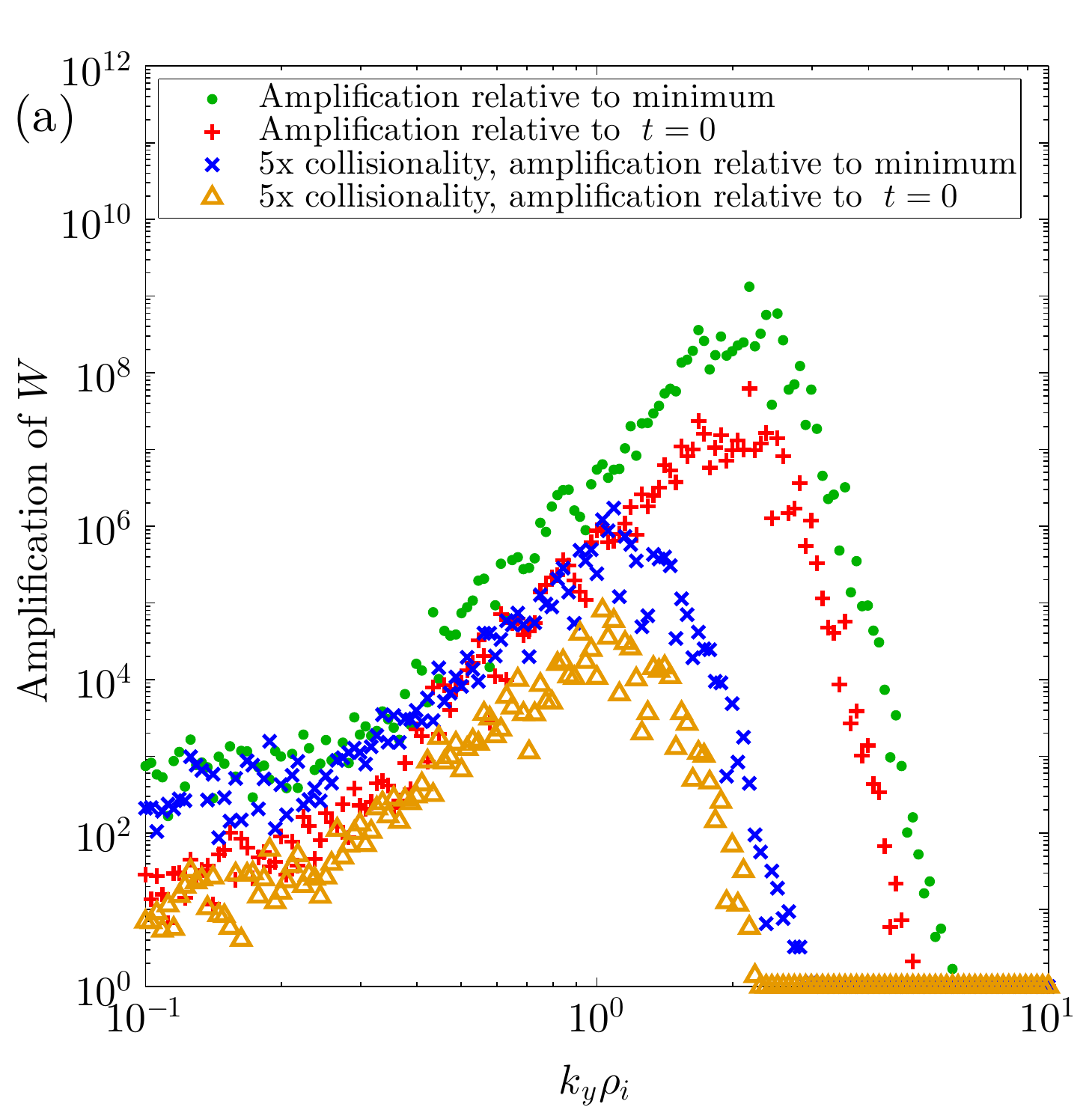}
\includegraphics[width=3in]{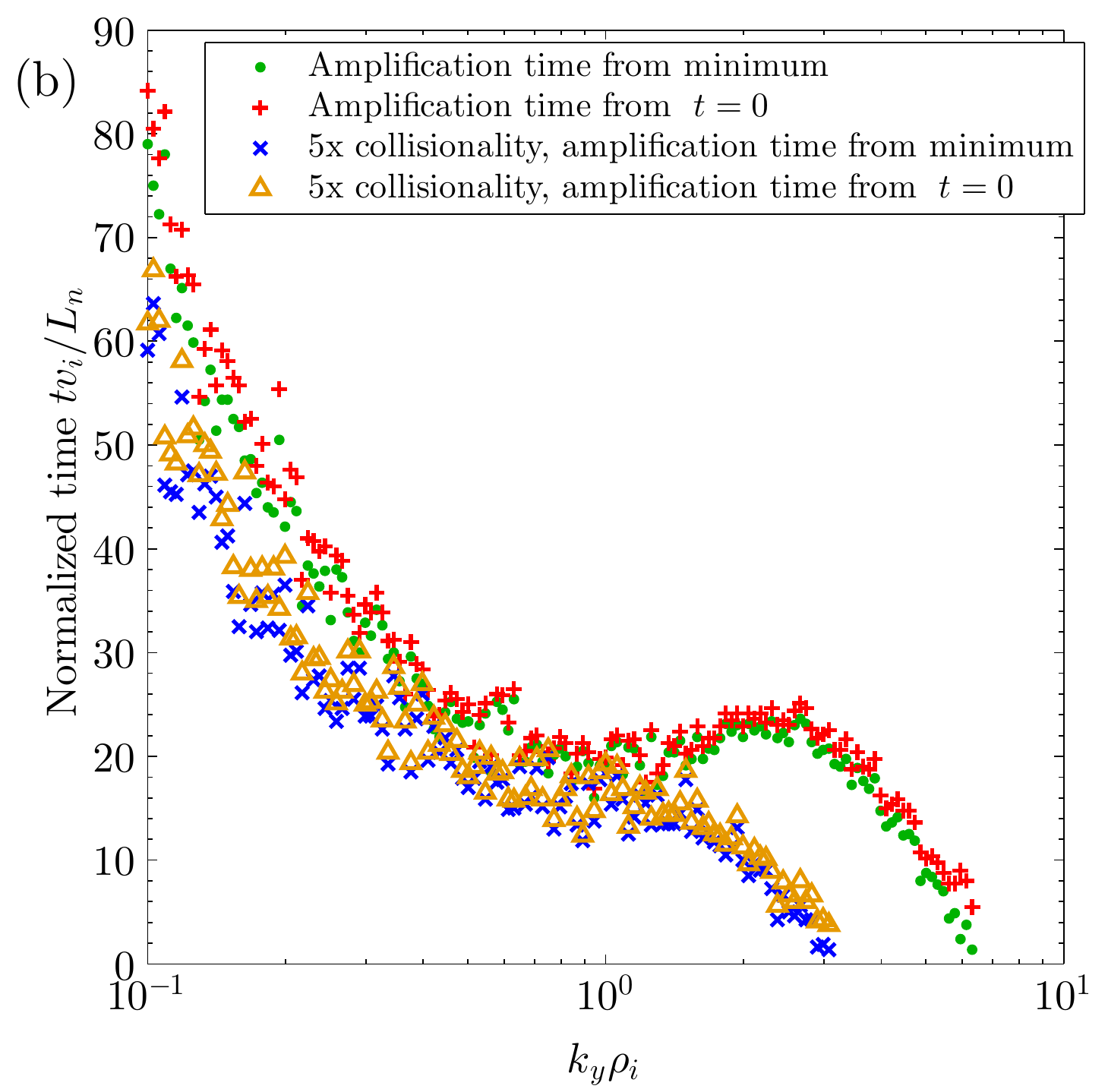}
\caption{(Color online) Transient amplification can be observed
for a range of $k_y$ surrounding $1/\rho_i$.  For this figure, $L_s/L_n=200$.
The two collisionalities shown are $\nuee L_n / v_i = 0.1$ and $0.5$.
\label{fig:transientAmplificationVsKy}}
\end{figure}

Figure \ref{fig:transientAmplificationVsKy} shows the dependence
of the amplification factor on $k_y$ at fixed $L_s/L_n = 200$.
Significant amplification can be observed for a range of $k_y$ surrounding
$1/\rho_i$.
Results are shown for two values of collisionality,
$\nuee L_n / v_i = 0.1$ and $0.5$.
(When the collisionality is much smaller than these values,
absolute linear instability is present \cite{usUniversalInstability}.)
It can be seen that increasing collisionality tends to limit amplification
at larger $k_y$ while having much less effect at smaller $k_y$.
This kinetic problem is similar to conventional hydrodynamics in that for both cases, non-modal amplification
decreases with increasing collisional dissipation (associated with viscosity in the Navier-Stokes equation
for the hydrodynamic case.)
However, the kinetic problem here is rather different
from the hydrodynamic case in the sensitivity of the transient amplification
to the geometric parameter $L_s/L_n$,
which is not associated with entropy-producing collisions.

\section{Turbulence}
\label{sec:turbulence}

\begin{figure}[h!]
\includegraphics[width=6.5in]{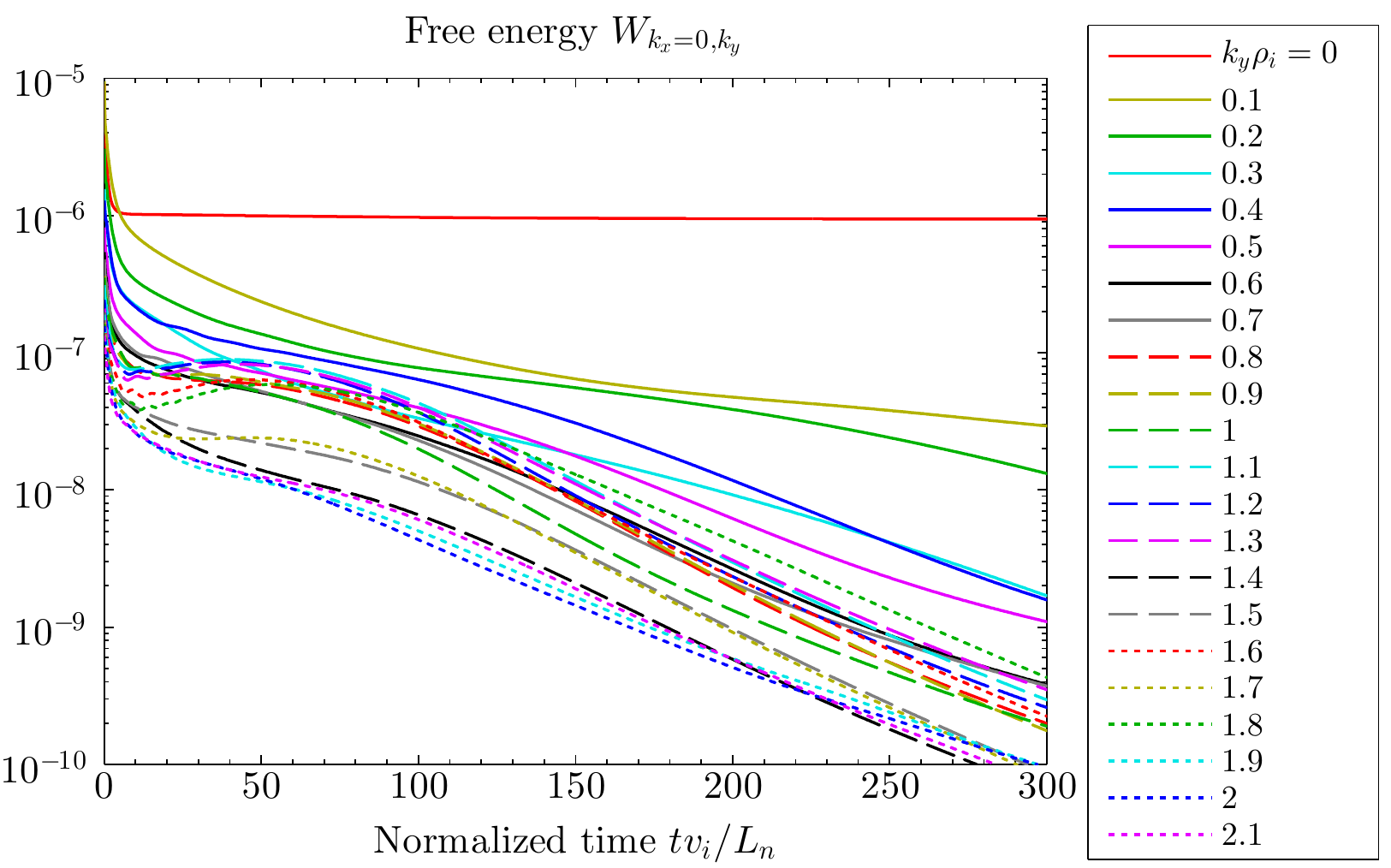}
\caption{(Color online)
Linear dynamics for the parameters used in the
nonlinear simulations of figure \ref{fig:amplitudeDependence}.
The large-$t$ behavior demonstrates there is no absolute linear instability.
A small amount of transient amplification can be seen.
\label{fig:linearBox}}
\end{figure}

We now demonstrate that when the nonlinear term in (\ref{eq:gke})
is retained, sustained turbulence is possible
in this system.
For the demonstration here we consider
the parameters $L_s/L_n = 20$, $T_e = T_i$, $m_i = 3672 m_e$, $\eta_i = \eta_e = 0$, and $\nuee L_n / v_i = 0.05$.
We find this point in parameter-space is approximately on the edge of where sustained turbulence
is possible: if the magnetic shear or collisionality is increased much beyond these values,
turbulence cannot be sustained indefinitely (though it may persist for a finite time.)
Note that $L_s/L_n=20$ is also approximately the maximum shear
in figure \ref{fig:transientAmplificationVsLsOverLn} for which transient linear amplification is observed
for random initial conditions.
(While absolute linear instability was found in \cite{usUniversalInstability} for $L_s/L_n \ge 17$ in the limit of vanishing collisionality, here the finite collisionality is enough to suppress this linear instability.)
Before presenting nonlinear simulations,
we first illustrate the linear behavior of the system using all the physical and numerical
parameters that will be used for the nonlinear runs.  Figure \ref{fig:linearBox} shows the time evolution of
$W$ for all the dealiased $k_y$ modes that will be
used for nonlinear simulations, including twist-and-shift boundary conditions,
but with the nonlinearity turned off, and
using the random initial conditions described in section \ref{sec:transient}.
More precisely, the quantities plotted are $W_{k_x,k_y}$, summed over the $k_x$ that link to $k_x=0$, where
\begin{equation}
\label{eq:Wk}
W_{k_x,k_y} = L_x L_y \sum_s \int dz \int d^3v \frac{T_s}{2 f_{Ms}}\left|\delta\! f_{s,k_x,k_y}\right|^2
\end{equation}
are the 2D Fourier contributions to $W = \sum_{k_x,k_y} W_{k_x,k_y}$.
Here,
$L_x$ and $L_y$ are the sizes of the simulation domain in $x$ and $y$,
and $\delta\! f_s =\sum_{k_x,k_y} \delta\! f_{s,k_x,k_y} \exp( i k_x x+i k_y y)$.
Curves for the other $k_x$ in the simulation look similar to those in figure \ref{fig:linearBox},
as expected from the $z$-translation symmetry of the system (\ref{eq:gke})-(\ref{eq:qn}); see discussion following (\ref{eq:kperp}).
The amplitude of each
of these curves decreases for large $t$, confirming there is no instability for these physical and
numerical parameters.  The decrease is mostly monotonic, but there is typically a small
amount of transient growth by a factor $ \lesssim 2\times$ for $k_y \rho_i$ between 1-2.
Thus, this point in parameter space is on the edge of the region in which transient linear
amplification occurs, as anticipated from figure \ref{fig:transientAmplificationVsLsOverLn}.

Throughout this section, the numerical parameters used were as follows: $12 \times 2$ grid points in pitch angle and $\sgn(v_{||})$, 12 grid points in energy, 24 grid points per cell in $z$ (with cells at $k_y=j \Delta k_y$ linked if separated in $k_x$ by $4 j \Delta k_x$), CFL number 0.5, $256 \times 64$ modes in $x$ and $y$ before dealiasing, and box size $(L_x, L_y, L_z) = (31.4 \rho_i, \; 62.8 \rho_i, \; 160 L_n)$.  It was verified that both the linear and nonlinear results reported in this section were not significantly altered under factor-of-2 changes in any of these numerical parameters.

One noteworthy linear property apparent in figure \ref{fig:linearBox}
is that damping of modes with $k_y=0$ is very weak,
nearly invisible on the scale of the figure.
Linear modes with $k_y=0$ and $\partial/\partial z=0$ are damped only by collisions.
(In the absence of any collisions, any $h_e$ and $h_i$ that depend only on $\energy_s$ and $\mu$ are steady-state solutions of the linearized system.)
Collisions introduce damping of these $k_y=0$ modes, with a damping rate that increases with $k_x^2$ due to the classical transport terms in the gyroaveraged collision operator.

\begin{figure}[h!]
\includegraphics[width=6.5in]{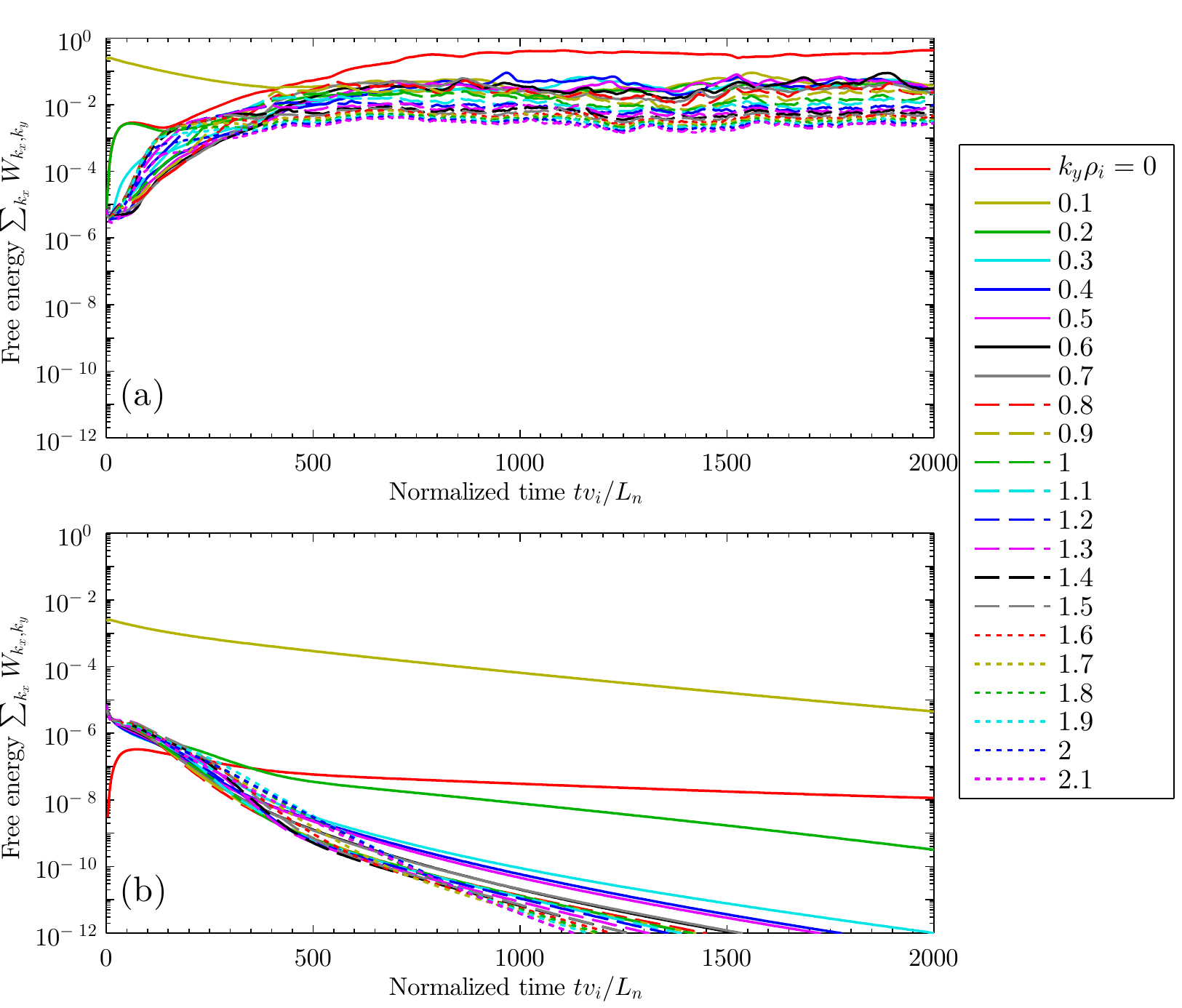}
\caption{(Color online)
Simulations for exactly the same physical and numerical parameters as the decaying
linear simulation in figure (\ref{fig:linearBox}), but now including nonlinearity and varying the
initial amplitude of the $k_y=0.1$ component.
(a) When the initial amplitude is sufficient, turbulence grows and is sustained, whereas (b) for small initial amplitude,
all modes decay as they do in a linear simulation.
\label{fig:amplitudeDependence}}
\end{figure}

Finally, figure \ref{fig:amplitudeDependence} shows two nonlinear simulations for the same parameters as figure \ref{fig:linearBox},
differing only in the initial amplitude of the $k_y=0.1$ part of the perturbation.
For sufficiently large initial amplitude, turbulence is sustained,
demonstrating nonlinear instability.

A variety of strategies can be used to initiate the nonlinear instability in numerical simulations.
Large-amplitude noise can be used, but this method is numerically inconvenient (at least in gs2) due to the rapid
drop in amplitude of strongly damped modes over the first few time steps,
which causes significant changes in the CFL-stable time step associated
with the perpendicular $\vect{E}\times\vect{B}$ speed,
thereby requiring recomputation of the implicit time-advance operator.
Another method for initiating the nonlinear instability is to begin with a sufficient
temperature gradient to bring about linear instability, and then to reduce the temperature gradient
once the turbulence reaches a sufficient amplitude \cite{Drake}.
A third effective method, the one used in figure \ref{fig:amplitudeDependence},
is to initialize with one of the least-damped linear eigenmodes scaled up to appropriate amplitude, with noise in the other modes.  We find this method to be particularly convenient since variation in the stable time step
is minimized.

\begin{figure}[h!]
\includegraphics[width=6.5in]{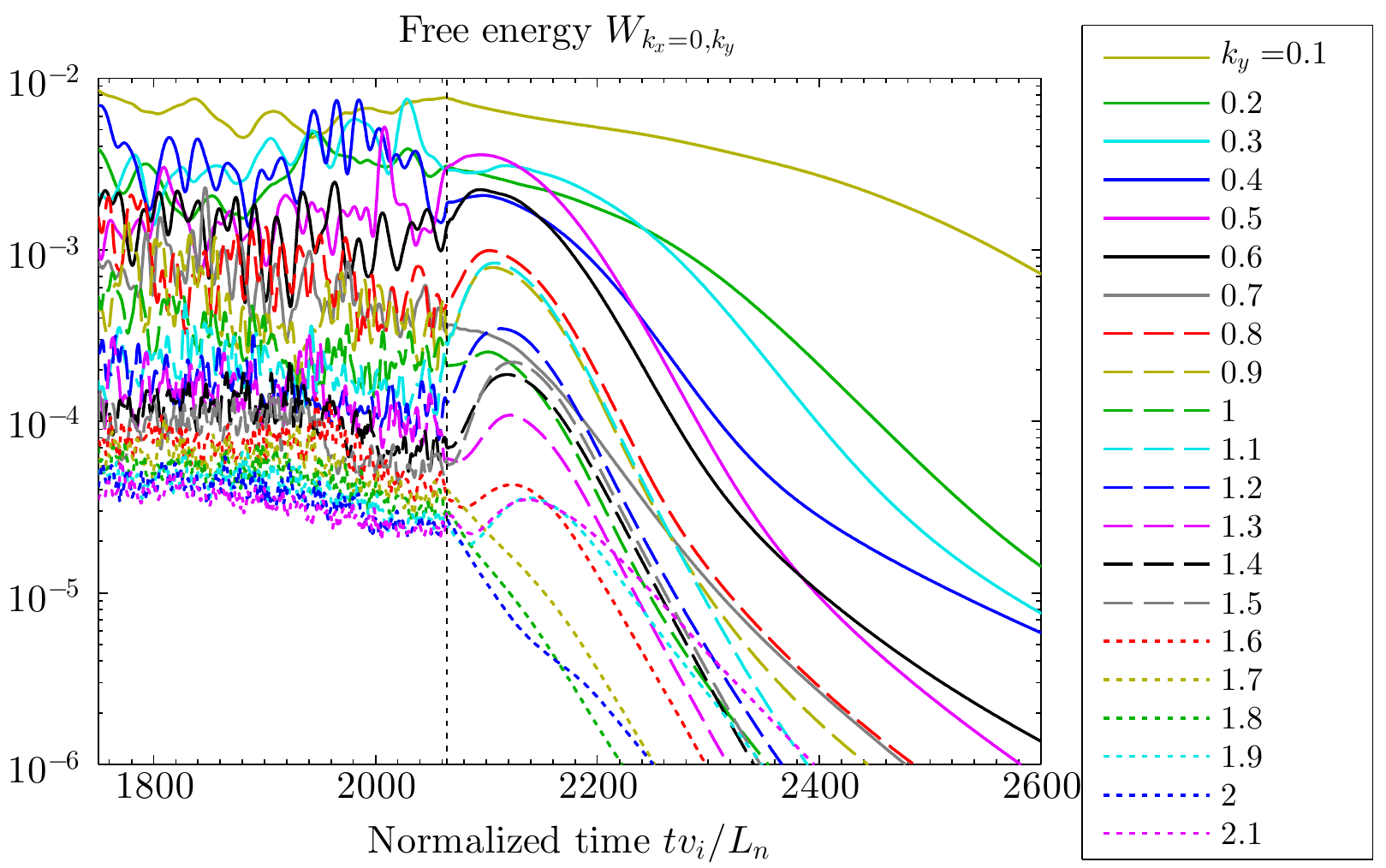}
\caption{(Color online)
Demonstration of the concepts in section \ref{sec:proof}.
Curves show free energy in the 2D modes with $k_x=0$ (including intervals linked by
the twist-and-shift parallel boundary condition) for various $k_y$.
After a period of saturated turbulence, the nonlinear term is turned off at normalized
time 2064, denoted by the vertical dashed line.
All modes eventually decay, but transient amplification is apparent first
for some of the 2D modes. Such amplification must occur in at least one 2D mode in such a linear ``restart'' from
steady turbulence, as proved in section \ref{sec:proof}.
\label{fig:linearRestart}}
\end{figure}

As another demonstration that the system displays nonlinear instability without linear instability,
figure \ref{fig:linearRestart} shows the result of
turning off the nonlinear term after a period of saturated turbulence.
The nonlinear term is turned off at time $t v_i / L_n = 2064$, shown as a vertical
dotted line.  All modes eventually decay.
However some modes
display a modest amount ($\lesssim 3\times$) of transient amplification before decaying.

\begin{figure}[h!]
\includegraphics[width=3in]{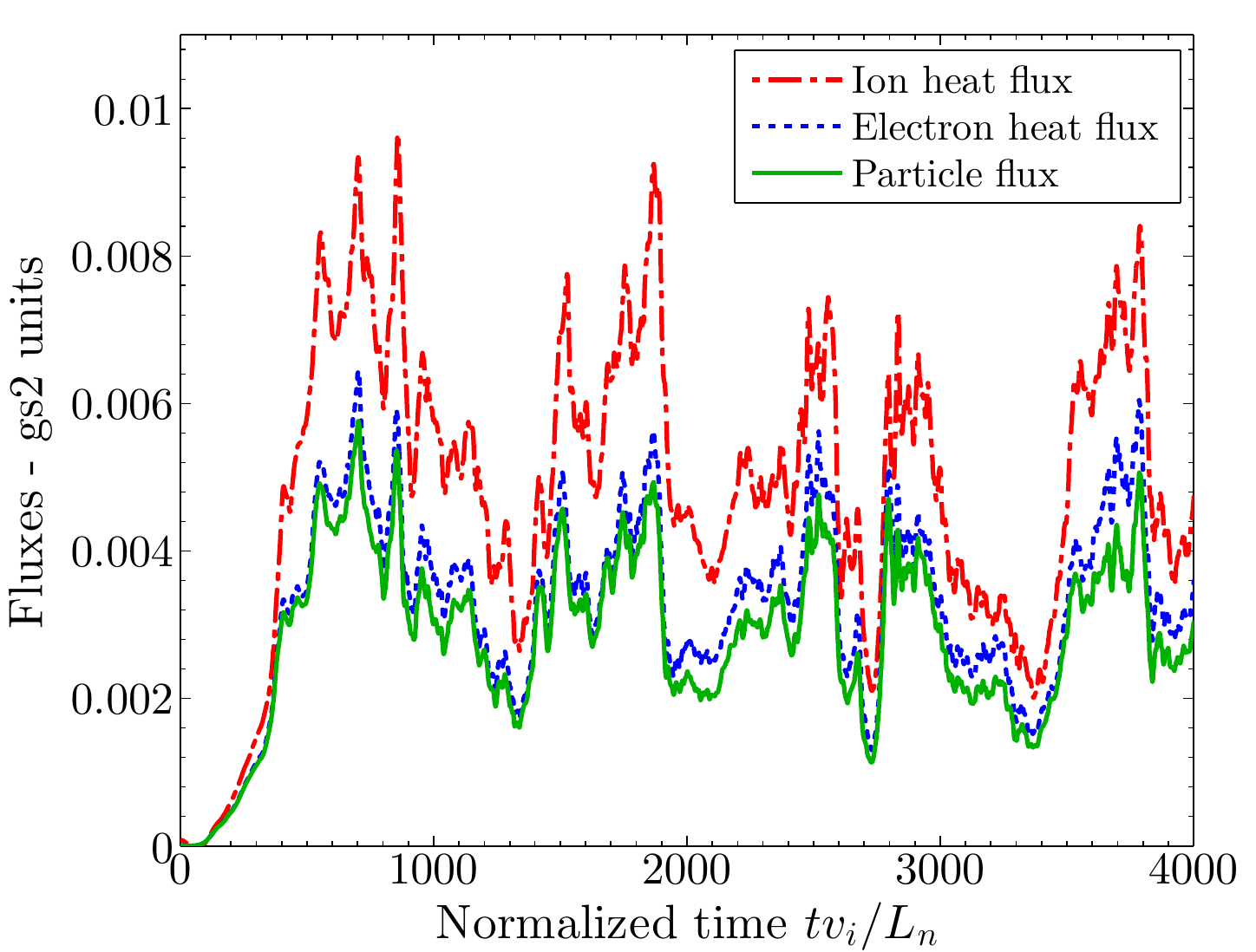}
\caption{(Color online)
Fluxes for the simulation of Figure (\ref{fig:amplitudeDependence}).a with  $\nuee L_n / v_i = 0.05$ and
$L_s/L_n=20$, approximately the maximum shear at which turbulence can be sustained at this collisionality.
Even in the absence of equilibrium temperature gradients, nonzero average ion and electron heat fluxes
are obtained.
\label{fig:fluxes}}
\end{figure}

The particle and heat fluxes for a long nonlinear simulation at these parameters are plotted in figure \ref{fig:fluxes}.
As the time average of $S$ in (\ref{eq:dWdt})-(\ref{eq:freeEnergySources}) must vanish, and since the collision term
is negative-definite, there must be an average particle flux in the appropriate direction
to relax the equilibrium density gradient, and indeed we find such a flux.
Although there is no equilibrium temperature gradient, nonzero average fluxes
of both ion and electron heat are found.  Both heat fluxes have the same sign
as the particle flux, meaning heat flows in the $-\nabla n$ direction.

The fluxes in figure \ref{fig:fluxes} are normalized by gyro-Bohm values defined using the length $L_n$ and including the 2 in the thermal speed, e.g.
the particle flux is normalized by $(\rho_i / L_n)^2 n v_i$.
Converting to dimensional units for a deuterium plasma, we obtain the following estimate for the turbulent particle diffusivity:
$D \approx \left( 0.4 \mathrm{ \;m^2/s}\right)
\left(T / 500 \mathrm{\;eV}\right)^{3/2}
\left(B/2\mathrm{\;T}\right)^{-2} \left( L_n / 1 \mathrm{\;cm}\right)^{-1}$.
This magnitude of diffusion coefficient may be large enough to be of interest experimentally.
A larger diffusion coefficient is obtained if $L_s/L_n$ is increased.

\section{Turbulence implies transient amplification of typical states}
\label{sec:proof}

It seems plausible that the previous two sections are related,
that a connection exists between transient linear amplification and sustained nonlinear turbulence.
Let us now prove this relationship rigorously.
Here we present two such arguments,
both of which yield the conclusion that non-modal amplification is a necessary condition
for sustained turbulence.  One argument for the case of hydrodynamics
was given in \cite{DelSoleNecessity} and section 4 of \cite{DelSoleSurvey},
and below we adapt this reasoning to gyrokinetics.  Before we do so, we give an alternative
argument that relies less on linear algebra.
For this discussion, it is useful to define a ``2D mode'' to be any structure with given $k_x$ and $k_y$,
but with no restriction on the $z$ dependence.

{\it Statement 1:} Consider a turbulent solution of the nonlinear gyrokinetic system (\ref{eq:gke})-(\ref{eq:qn}),
i.e. a solution for which $W$ does not decay to zero at large $t$.
Choose any time $t_0$ at which the free energy satisfies $dW/dt>0$.
Suppose at $t_0$ we turn off the nonlinear term
and re-evolve the system in time, as in figure \ref{fig:linearRestart}.
Then at least one 2D mode must grow in the $W$ norm
for some time following $t_0$, i.e. there must be transient linear amplification.

{\it Proof:}
Recall that (\ref{eq:dWdt}) is a moment of both the nonlinear and linear dynamical equations.
When the nonlinear term is turned off at $t_0$, $S(h_i,h_e)$ does not immediately change, because
the system state does not immediately change. Therefore, $dW/dt$ must vary \emph{continuously}
when the nonlinear term is turned off, and so $dW/dt$ is positive for some period following $t_0$ under the
action of the linear dynamics.
Recall $W$ is a sum of positive contributions (\ref{eq:Wk}) from each 2D mode:
$W = \sum_{k_x,k_y} W_{k_x,k_y}$.
Therefore $dW_{k_x,k_y}/dt$ must be positive in the linear dynamics for at least one 2D mode,
so this mode must be growing transiently in the $W$ norm.

Roughly speaking, the upshot of this argument is that we should not be surprised that there is transient amplification
after the nonlinear term is turned off in figure (\ref{fig:linearRestart}).
Such amplification \emph{must} occur.
Put another way, a necessary condition
for turbulence (in the absence of linear instability) is that typical states of the turbulence
can be transiently amplified. Furthermore, the energy growth in the 2D modes that are amplified
must be sufficient to balance the energy decay from the damped 2D modes.

Next, we present a second argument for the necessity of transient amplification.
This argument is based on linear algebra and closely follows \cite{DelSoleNecessity}.
For this second approach, we let $\vect{h}$ denote
a column vector that specifies the state of the system $\{h_i,h_e\}$ in some convenient basis.
For example, $\vect{h}$ could represent the array of values specifying the distribution function in a code.

{\it Statement 2:} Given a set of physical parameters for which
the nonlinear gyrokinetic system (\ref{eq:gke})-(\ref{eq:qn})
permits statistically steady turbulence,
there exists at least one state $\vect{h}_0$ such that
(1) $dW/dt>0$ if $\vect{h} = \vect{h}_0$, and
(2) the squared projection of the turbulent state $\vect{h}(t)$
onto $\vect{h}_0$, defined below, has a non-zero average.

{\it Proof:}
Let $\vect{L}$ denote the time-independent linear operator associated with (\ref{eq:gke})-(\ref{eq:qn})
in the absence of the nonlinear term,
so the linearized (\ref{eq:gke})-(\ref{eq:qn}) can be written
\begin{equation}
d\vect{h}/dt = \vect{L}\vect{h}.
\end{equation}
(Here we will take $\vect{L}$ to be real since the equations (\ref{eq:gke})-(\ref{eq:qn})
are real, though the following argument may be generalized to complex
$\vect{L}$ by replacing the transpose operation with Hermitian conjugation.)
Also let $\vect{E}$ be the time-independent operator that defines the free energy norm:
\begin{equation}
W = \vect{h}^{\transpose} \vect{E} \vect{h},
\label{eq:hEh}
\end{equation}
where $^\transpose$ denotes the transpose. Note that $\vect{E}$ is real and symmetric
($\vect{E} = \vect{E}^\transpose$). Differentiating (\ref{eq:hEh}),
\begin{equation}
\label{eq:dWdt2}
dW/dt = \vect{h}^{\transpose} \left[ (\vect{E}\vect{L})^\transpose + \vect{E}\vect{L} \right] \vect{h},
\end{equation}
which is equivalent to (\ref{eq:dWdt}). As discussed previously, the presence of the nonlinear term
does not alter (\ref{eq:dWdt}) (since the nonlinear term is annihilated in the integration over
the perpendicular directions $X_s$ and $Y_s$),
so (\ref{eq:dWdt2}) holds for both the linear and nonlinear gyrokinetic equations.
Introducing a time average $\left< \ldots \right>$ over statistically steady turbulence, then
\begin{equation}
\label{eq:avg}
\left< \vect{h}^{\transpose} \left[ (\vect{E}\vect{L})^\transpose + \vect{E}\vect{L} \right] \vect{h} \right>
=0.
\end{equation}

Next, observe that the operator $(\vect{E}\vect{L})^\transpose + \vect{E}\vect{L}$ in (\ref{eq:dWdt2})-(\ref{eq:avg})
is real and symmetric, so it possesses a complete set of orthogonal eigenvectors with real eigenvalues.
Let $\vect{u}_j$ and $\lambda_j$
denote these eigenvectors (normalized so $\vect{u}_j^\transpose \vect{u}_k = \delta_{j,k}$)
and their associated eigenvalues.
In the hydrodynamics literature, the eigenvectors $\vect{u}_k$ are sometimes called
instantaneous optimals; the $\vect{u}_j$ with maximum $\lambda_j$ is the fluctuation
that maximizes the instantaneous $dW/dt$.
We may decompose the state vector in the $\vect{u}_j$ basis:
$\vect{h}(t) = \sum_j a_j(t) \vect{u}_j$ for some amplitudes $a_j(t)$.
Using this decomposition in (\ref{eq:avg}) gives
\begin{equation}
\label{eq:sum}
\sum_j \lambda_j \left< \left| a_j(t) \right|^2 \right>
=0.
\end{equation}
Neglecting the uninteresting case in which all terms in the sum (\ref{eq:sum}) are zero,
then there must be at least one positive eigenvalue $\lambda_{j_0}$ with an associated nonzero
average weight $\left< \left| a_{j_0}(t) \right|^2 \right>$. Considering the associated
$\vect{u}_{j_0}$ in (\ref{eq:dWdt2}), then this state $\vect{u}_{j_0}$ must give positive $dW/dt$,
proving statement 2.

Thus, we have showed in two ways that transient linear amplification is a necessary
condition for turbulence in the associated nonlinear system.  We have also shown that the amplitude of the
transiently growing state(s) must have a significant amplitude in the turbulent state, in the following senses.
In the first line of argument, not only must there be at least a single growing 2D mode,
but there must be enough growing 2D modes of sufficient amplitude to balance the negative $dW_{k_x,k_y}/dt$
of all the decaying 2D modes.
In the second argument, in (\ref{eq:sum})
there must not only be at least one state $j$ with an instantaneous growth rate
$\lambda_j >0$, but the amplitudes of states with positive and negative
$\lambda_j$ must be in balance so $\left< dW/dt \right>=0$. In a statistically steady state, there must be a balance between energy input from the instantaneously growing
linear states and the instantaneously decaying linear states.

Both of the arguments above relied on the fact that the norm $W$ satisfies the same equation
(\ref{eq:dWdt}) in both the linear and nonlinear dynamics.
Thus, the ``nonlinear invariant'' $W$ is a special norm, and the arguments do not generally apply to other norm-like quantities
such as $\int d^3r \left| \Phi \right|^2$ in which one might choose to measure transient amplification.
(In fact for any operator with damped eigenmodes,
even if transient amplification exists in one norm,
another norm always exists in which transient amplification is impossible.
The sum of squared eigenmode coefficients is such a norm \cite{FarrellIoannou1993}.)

While the arguments here relied on the gyrokinetic conservation law (\ref{eq:dWdt}), the same reasoning may be applied to
other models and systems in which an energy conservation equation holds. For example,
eq (6) in Ref. \cite{Drake} gives the appropriate conservation law for the fluid model
in that work, and the first line of (6) gives the associated norm to use in place of $W$.

\section{Discussion and conclusions}
\label{sec:conclusions}

The ``universal mode'' system (slab geometry with a density gradient, electrostatic,
no sheared equilibrium flow, and weak temperature gradients)
is known to have a dramatic change in linearly stability
as the magnetic shear is raised from zero to small finite values
\cite{Ross, Tsang, Antonsen,usUniversalInstability}.
As the eigenvalues are so sensitive to magnetic shear,
we might expect traditional modal analysis to
give a misleading picture of typical behavior,
as with other systems that have sensitive eigenvalues \cite{TrefethenEmbree}.
Indeed, here we have demonstrated that significant
transient linear amplification is possible in the drift wave system even when
all eigenmodes are decaying.
As shown in figure \ref{fig:comparingTransientTo0Shear},
the transient structures grow at the zero-shear
growth rate, when evaluated at the dominant $k_{\perp}$ and $k_{||}$ at that instant.
We have demonstrated that the amount and duration of amplification scale
with the magnetic shear length $L_s$, so these quantities naturally become
unbounded in the limit of vanishing shear.
The non-modal amplification also decreases with increasing collisionality,
particularly at the shortest perpendicular wavelengths,
and the amplification tends to be largest at wavelengths near $k_y \rho_i \sim 1-2$.
In contrast to more famous examples of non-modal amplification
from hydrodynamics \cite{TrefethenSubcritical}, and to some recent examples from plasma physics \cite{Newton, Alex, BarnesRotation, EdmundPRL},
the amplification in the plasma drift wave system here
does not depend on sheared equilibrium flow.

We have also demonstrated using nonlinear gyrokinetic simulations
that sustained turbulence is possible in this system
even when all linear modes are damped.
This phenomenon had been seen previously
in fluid models \cite{Scott1, Scott2, Drake}.
This phenomenon, clearly visible in figure \ref{fig:amplitudeDependence},
is opposite to the ``Dimits shift'' \cite{Dimits} seen at different physical parameters,
in which nonlinear fluctuations
are suppressed despite the presence of linear instability.

In this system without unstable eigenvalues, we find that sustained nonlinear turbulence
and transient linear amplification both become possible around $L_s / L_n \gtrsim 20$
(for the collisionality $\nuee L_n / v_i = 0.05$ and other parameters considered.)
As discussed in section \ref{sec:proof}, the fact that these two phenomena appear together is not
a coincidence. Transient linear amplification in the free energy norm $W$ is a necessary condition for statistically steady
turbulence, and the turbulent state must have a significant projection onto states that are linearly growing in the $W$
norm.  When the nonlinear term is turned off in a simulation of turbulence, as in figure \ref{fig:linearRestart},
amplification in $W$ norm must occur in at least one of the 2D modes.
The amplitude of the instantaneously growing states (instantaneous optimals) must be
significant, enough for the associated input of free energy to balance decay of energy
from the states that are decaying.

Considering the nonlinear and linear results together, the existence of transient amplification and
turbulence evidently depends on whether magnetic shear is sufficiently weak that un-sheared slab mode
dynamics can persist.
This geometric property can be controlled independently of the energy source of the turbulence (i.e. the density gradient),
in contrast to turbulence driven by sheared flows where the nonuniformity of the flow plays two roles at once:
shaping the global structure of the growing structures and providing energy to the turbulence.
This feature makes the present system an interesting case study of sub-critical turbulence, and we are unaware of other such examples.”

Note that the transient linear process emphasized here and secondary instability are not mutually
exclusive. Both are required to support turbulence.
Some form of secondary instability (as discussed in \cite{Drake}) must be present to transfer energy out of the transiently growing linear
structures before these structures begin to decay substantially. However, nonlinear processes cannot
directly contribute to the system's free energy $W$, due to the annihilation of the nonlinear term
in deriving (\ref{eq:dWdt}).

Several effects that may be important for realistic plasma turbulence
were not included in the present work, including toroidicity and trapped particles,
and electromagnetic fluctuations.
Therefore, further work is required to determine how relevant the nonlinear turbulence considered here
is for realistic experimental parameters.  However, we at least conclude that linear stability
need not preclude the possibility of significant turbulent particle and energy fluxes,
and that nonlinear simulations should sometimes be initialized at finite amplitude to examine this possibility.

\begin{acknowledgments}
This material is based upon work supported by the
U.S. Department of Energy, Office of Science, Office of Fusion Energy Science,
under Award Numbers DEFG0293ER54197 and DEFC0208ER54964.
Computations were performed on the Edison system at
the National Energy Research Scientific Computing Center, a DOE Office of Science User Facility supported by the Office of Science of the U.S. Department of Energy under Contract No. DE-AC02-05CH11231.
Conversations with
Tom Antonsen, Michael Barnes, Greg Colyer, James Drake, Edmund Highcock, Greg Hammett, and Adil Hassam
are gratefully acknowledged.
\end{acknowledgments}

\appendix

\section{Linear stability in the absence of magnetic shear}
\label{appendix:zeroShear}

\begin{figure}[h!]
\includegraphics[width=6in]{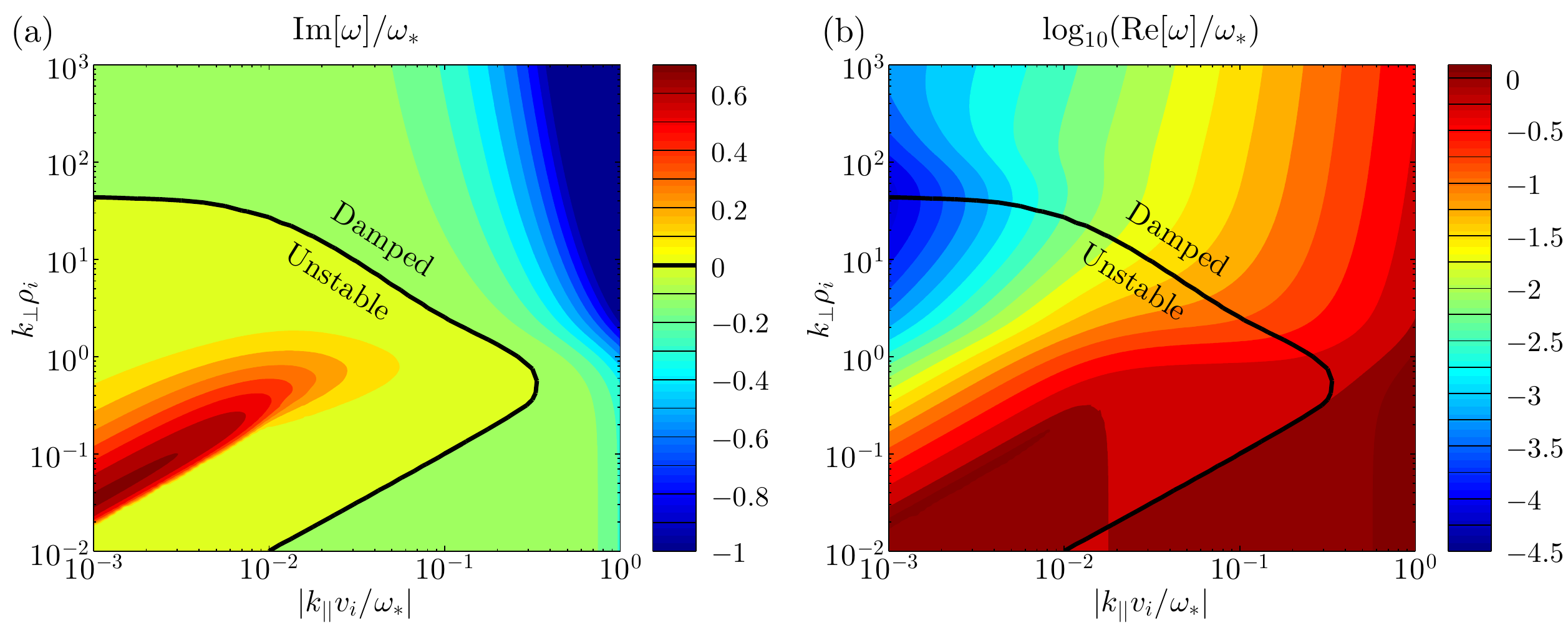}
\caption{(Color online) Solution of the linear dispersion relation in the zero-shear (local) limit, i.e. the
root of (\ref{eq:dispersionRelation}) with largest imaginary part, for $\eta_i = \eta_e = 0$,
$T_e = T_i$, and $m_i = 3672 m_e$.
The (a) imaginary and (b)
real parts of the frequency are shown.
\label{fig:localDispersionRelation}}
\end{figure}

For reference, here we present the growth/damping rates and real frequencies of the
universal instability in the limit of vanishing
magnetic shear and vanishing collisions.
In the absence of shear, we can replace $\partial/\partial z$ in (\ref{eq:gke})
with $i k_{||}$ for some constant $k_{||}$.
Using (\ref{eq:qn}), and after some algebra, the local dispersion relation may then be written
\begin{eqnarray}
\frac{T_e}{T_i}-\zeta_i\frac{\omegase}{\omega}\left\{
-\left[\frac{\omega T_e}{\omegase T_i}+1\right]Z_i \Gamma_{0i}
+\eta_i Z_i \left[ \left(\frac{1}{2}+b_i\right) \Gamma_{0i}-b_i \Gamma_{1i}\right]
-\eta_i \zeta_i \left[1+\zeta_i Z_i\right] \Gamma_{0i}\right\}
&&\label{eq:dispersionRelation} \\
+
1+\zeta_e\frac{\omegase}{\omega}\left\{
\left[\frac{\omega}{\omegase}-1\right]Z_e \Gamma_{0e}
+\eta_e Z_e \left[ \left(\frac{1}{2}+b_e\right) \Gamma_{0e}-b_e \Gamma_{1e}\right]
-\eta_e \zeta_e \left[1+\zeta_e Z_e\right] \Gamma_{0e}\right\}
&=&0.
\nonumber
\end{eqnarray}
Here, $\zeta_s = \omega/(\left| k_{||} \right| v_s)$, $v_s = \sqrt{2 T_s/m_s}$ is the thermal speed, $Z_s = Z(\zeta_s)$, $Z$ is the plasma dispersion function,
$\Gamma_{js} = I_j(b_s)e^{-b_s}$, $I_j$ is a modified Bessel function, $b_s = k_{\perp}^2 v_s^2/(2 \Omega_s^2)$,
and $\omega_* = k_y T_e/(e B L_n)$.
Figure \ref{fig:localDispersionRelation} shows the solution of (\ref{eq:dispersionRelation})
with largest imaginary part for the case $T_e = T_i$
and the deuterium-electron mass ratio.
We take $\eta_i = \eta_e = 0$
so it is certain that the ion-temperature-gradient (ITG) and electron-temperature-gradient (ETG)
modes are suppressed.
Figure \ref{fig:localDispersionRelation} was generated by applying nonlinear root-finding to (\ref{eq:dispersionRelation}),
verifying the solution agreed with initial-value gs2 simulations for several values of $k_{||}$ and $k_{\perp}$.
It can be seen that absolute linear instability exists for sufficiently small values of $\left|k_{||}\right|$
when $k_{\perp} \rho_i < 40$.

Some insight into the density-gradient-driven instability can be obtained by
examining the limit $|\zeta_e| \ll 1$ so $Z_e \approx i\sqrt{\pi}$,
and $|\zeta_i| \gg 1$ so $Z_i \approx -1/\zeta_i$. Considering solutions with $|\omega/\omega_*| \ll 1$,
one finds
\begin{equation}
\label{eq:approxRoot}
\omega = \frac{\omega_* \Gamma_{0i}}
{\left(\frac{T_e}{T_i}+1\right)^2 + \left(\sqrt{\pi}\frac{\omega_*}{|k_{||}| v_e}\right)^2}
\left( \frac{T_e}{T_i} + 1 + i \sqrt{\pi}\frac{\omega_*}{|k_{||}| v_e}\right).
\end{equation}
The imaginary part of $\omega$, proportional to  $Z_e \approx i\sqrt{\pi}$, evidently arises due to the electron resonance.


\end{document}